\documentclass[prl,aps,amsmath,amssymb,twocolumn,longbibliography]{revtex4-1}
\usepackage{graphicx,multirow}
\usepackage{color,outlines}
\usepackage[sc,osf]{mathpazo}

\usepackage{physics}
\usepackage{outlines}
\usepackage{tikz}

\usepackage{hyperref}
\usepackage{graphicx}
\usepackage{dcolumn}
\usepackage{bm}
\usepackage{color}
\usepackage{xcolor}
\usepackage{bbold}
\usepackage{soul}
\usepackage{epsfig,epic}
\usepackage{epstopdf}
\usepackage{amsthm}
\usepackage{cleveref}

\newcommand{\newc}{\newcommand}
\newc{\beq}{\begin{equation}}
\newc{\eeq}{\end{equation}}
\newc{\kt}{\rangle}
\newc{\br}{\langle}
\newc{\beqa}{\begin{eqnarray}}
\newc{\eeqa}{\end{eqnarray}}
\newc{\ovl}{\overline}
\usepackage[flushleft]{threeparttable} 
\usepackage{tabularx}
\usepackage{array}
\usepackage{soul}

\usepackage{tikz}
\tikzset{snake it/.style={decorate, decoration=snake}}
\usetikzlibrary{shapes.geometric, arrows}
\usetikzlibrary{decorations.pathmorphing}

\begin{document}

\title{Creating ensembles of dual unitary and maximally entangling quantum evolutions} 
\author{Suhail Ahmad Rather}
\email{suhailmushtaq@physics.iitm.ac.in}
\affiliation{Department of Physics, Indian Institute of Technology
Madras, Chennai, India~600036}
\author{S. Aravinda}
\email{aravinda@physics.iitm.ac.in}
\affiliation{Department of Physics, Indian Institute of Technology
Madras, Chennai, India~600036}
\author{Arul
Lakshminarayan}
\email{arul@physics.iitm.ac.in} 
\affiliation{Department of Physics, Indian Institute of Technology
Madras, Chennai, India~600036}
   
\begin{abstract}
Maximally entangled bipartite unitary operators or gates find various applications from quantum information to many-body physics wherein they are building blocks of minimal models of quantum chaos. In the latter case, they are referred to as ``dual unitaries". Dual unitary operators that can create the maximum average entanglement when acting on product states have to satisfy additional constraints. These have been called ``2-unitaries" and are examples of perfect tensors that can be used to construct absolutely maximally entangled states of four parties. Hitherto, no systematic method exists in any local dimension, which results in the formation of such special classes of unitary operators. We outline an iterative protocol, a nonlinear map on the space of unitary operators, that creates ensembles whose members are arbitrarily close to being dual unitaries. For qutrits and ququads we find that a slightly modified protocol yields a plethora of 2-unitaries.  
\end{abstract}

\maketitle

Entanglement of states has been appreciated since almost the inception of quantum mechanics as its quintessential property \cite{EPR1935,schrodinger1935discussion}  and implies that the whole system could be in a definite state while the parts are not. In the extensive ongoing studies of quantum information it has acquired the central status of a resource \cite{Horo2009}. Operators, as quantum gates, observables or time-evolution propagators are also central to quantum mechanics. Unentangled states are often entangled due to the action of entangling unitary operators in the circuit paradigm of quantum computing \cite{NielsenChuang}. Thus how entangled unitary operators themselves are (measured by {\sl operator entanglement}) \cite{Zanardi2001}, and how much entanglement they can produce, on the average, acting on unentangled states (measured by {\sl entangling power}) \cite{Zanardi2000} are of primary interest. They have also started forming a means to characterize complexity in many-body systems \cite{Dubail17,Luitz2017,PalLak2018}, and early applications include quantum transport in light-harvesting complexes \cite{Plen10} and the study of quantum chaos \cite{DD04,Kestner18}. They are state-independent measures of entanglement growth, including an approach to thermalization \cite{Bhargavi2017,Bhargavi2019}.  Other studies of operator nonlocality as a resource include\cite{Collins2001,Vidal2002,Hammerer2002,Nielsen2003}, while the works  \cite{Kus2013,Mandarino2018} study entangling power of unitary gates and their powers in the case of qubits.

Maximally entangled bipartite states such as the prototypical Bell states of two qubits, and its generalization to any dimension: $\sum_{i=1}^d |i_Ai_B\kt/\sqrt{d}$ (where $|i_{A,B}\kt$ form a complete orthonormal basis in each of the particle spaces) is straightforward to construct and characterize.
On the contrary, while it is easy to state conditions under which an operator may have maximal operator entanglement or entangling power,
it seems surprisingly difficult to construct their Bell state equivalents, let alone to characterize and parameterize them \cite{Goyeneche2015}. One exception is the class of permutation matrices
based on orthogonal Latin squares which provide a finite set of maximally entangling unitary operators for any local dimension other than $2$ and $6$ \cite{Clarisse2005}. 

Further motivation for constructing maximally entangled unitary operators comes from recent observations concerning lattice models wherein
a space-time duality allows for some analytical results, even for nonintegrable systems \cite{Akila2016,Bertini2019,piroli2019exact,bertini2019operator}.
More explicitly, using a ``dual unitary" \cite{Bertini2019}, as the nearest neighbor interaction in many-body systems, leads to solvable correlation functions. 
It is not hard to see that these dual unitary operators have maximal operator entanglement \cite{Zyczkowski2004,Bengtsson2007},
and we will refer to such operators simply as dual unitaries.

A subset of dual-unitaries also maximize the entangling power and have been called 2-unitaries \cite{Goyeneche2015}.
An operator-state isomorphism (see \cite{SM}) implies that 2-unitaries can be used to generate absolutely maximally entangled states (AME) \cite{HCLRL12,Goyeneche2015} of four parties, wherein all bipartitions are maximally entangled. 
The 2-unitaries are also ``perfect tensors'' of rank-4, ingredients of holographic quantum states and codes \cite{Pastawski2015}.
If the dual unitaries in the lattice models are in fact 2-unitaries, it is easy to see based on the results of \cite{Bertini2019}
that they lead to models with the fastest decay of correlations and are in this sense maximally chaotic. 
The usage of multiple terminologies stem from their varied origins. We stick to calling  dual unitaries those  which maximize operator entanglement, and 2-unitaries are those dual unitaries that also maximize entangling power.

In this Letter, we outline a protocol that leads iteratively to a systematic increase of the operator entanglement, leading to operators that are arbitrarily close to being dual unitaries.  A subset of these could also be  arbitrarily close to  2-unitaries , but we outline an alternative strategy that, while not monotonic, leads to near  2-unitaries.  In particular, for the case of local dimension $d=3$ (qutrits) and to some extent $d=4$, we show that the procedure leads to a considerable measure of  2-unitaries.   It may be noted that for $d=2$, the qubit case, 2-unitaries do not exist \cite{Higuchi2000,Zanardi2001}. Starting from random unitary matrices selected uniformly from the group $U(d^2)$, the circular unitary ensemble or CUE of random matrix theory (RMT) \cite{Dys62,Mehta}, these protocols generate an ensemble of dual unitaries for all local dimensions and an ensemble of 2-unitaries for $d = 3$ and $d = 4$.

\emph{Definitions and preliminaries:} Consider the bipartite Hilbert space $\mathcal{H}^d_A \otimes\mathcal{H}^d_B$  and let $\mathcal{U}(d^2)$ be the set of unitary operators in it. If $U \in \mathcal{U}(d^2)$, it's operator  Schmidt decomposition is given by 
$U=\sum_{j=1}^{d^2} \sqrt{\lambda_j}\, m^A_j \otimes m^B_j$, where $\lambda_1 \geq \cdots \geq \lambda_{d^2}\geq 0$, and  $\{m^{A}_j\}$ and $\{m^B_j\}$ form local orthonormal  operator bases  under the Hilbert-Schmidt inner product \cite{Nielsen2003,Zyczkowski2004}, {\it i.e.}, $\tr(m^{A}_j m^{A \,\dagger}_k)=\tr(m^{B}_j m^{B \,\dagger}_k)=\delta_{jk}$. The unitarity of  $U$ implies that $\sum_{j=1}^{d^2}\lambda_j/d^2=1$, as $\tr(UU^\dagger) = d^2$.  Hence defining probabilities $p_j=\lambda_j/d^2$, 
the Tsallis-entropies 
\beq
\label{eq:Tsallis}
S_q(U)=\frac{1-\sum_{j} p_j^q}{q-1}
\eeq
 are measures of operator entanglement. 
The entropy $S_q(U)=0$ iff the operator is of product form, when $\lambda_1=d^2$ and the rest vanish.
The $q=2$ case, $S_2(U)$ is also the linear entropy $E(U)=1-\sum_{j=1}^{d^2}\lambda_j^2/d^4$, referred to simply as the {\it operator entanglement} below. The case of $q=1/2$ will turn out to be important as well.  For any $q$, $S_q(U)$ is maximized when all $\lambda_j=1$, and in particular $E(U)_{\text{max}}=1-1/d^2$.  However, to construct such  dual unitary operators   the constraints required of the $2d^2$ operators $m_j^{A,B}$ are difficult to satisfy. Hence while there are known operators such as {\sc swap}, denoted below as $S$, ( $S\ket{\phi_A}\ket{\phi_B} = \ket{\phi_B}\ket{\phi_A}$ for all product states $|\phi_A\kt|\phi_B\kt$)  and the Fourier transform \cite{Tyson2003,Bhargavi2019} in arbitrary dimensions that are dual operators, systematic constructions, with the exception of qubits \cite{Bertini2019}, are lacking.

If $\br n m | U^R| \alpha \beta \kt= \br n \alpha |U| m \beta \kt$ is the realignment or reshaping, it is easy to see that $(X\otimes Y)^R=|X\kt \br Y^*|$, where $|X\kt$ is the row-vectorization of the matrix $X$ and $^*$ is the complex conjugation. It then follows from the 
Schmidt decomposition of $U$ that $U^R=\sum_{j=1}^{d^2} \sqrt{\lambda_j} |m_j^A\kt \br m_j^{B*}|$, the spectral decomposition of 
$U^R U^{R \, \dagger}=\sum_{j=1}^{d^2}\lambda_j|m_j^A\kt \br m_j^A|$ and $E(U)=1-\tr[(U^R U^{R \dagger})^2]/d^4$.
Hence iff $U^R$ is also unitary all the eigenvalues $\lambda_j=1$ and $U$ is dual unitary. The operator entanglement has a well-known operational meaning 
that for completeness is recalled in \cite{SM}. 
\begin{figure}[htbp]
\includegraphics[scale=.26]{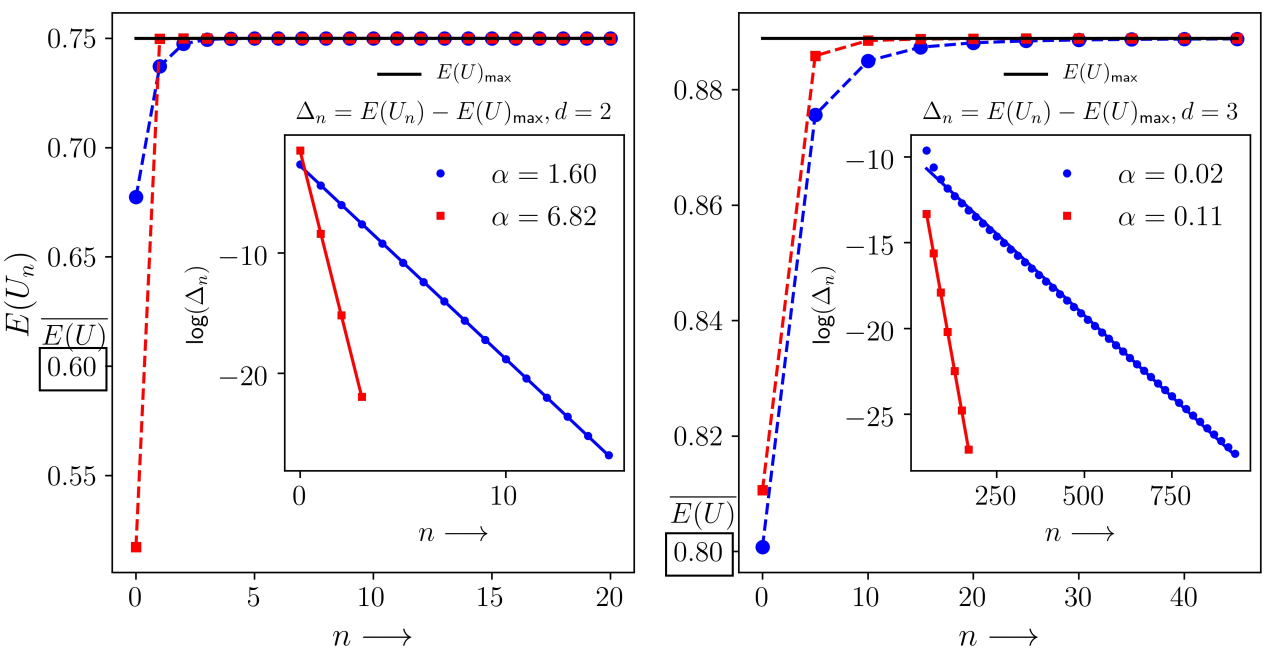}
\caption{Evolution of the operator entanglement $E(U_n)$ under the $M_R$ map for local dimensions $d = 2$ (left panel) and $d= 3$ (right panel) for two random initial operators $U_0$ in each case.
The $E(U_n)$ saturates to the corresponding maximum possible value $(d^2-1)/d^2$, clearly distinguishable from the average value, $\overline{E(U)}=(d^2-1)/(d^2+1)$ indicated in the boxes (see \cite{Zanardi2001} for a derivation of the average value).}
\label{fig:EUvsn}
\end{figure}

\emph{The realignment-nearest-unitary map $M_R$}:  Two stages define the map $M_R:U(d^2) \mapsto U(d^2)$, first is the linear one $U \mapsto U^R$, and the second is the nonlinear one that maps $U^R \mapsto V$ where $V$ is the nearest unitary operator to $U^R$. Let $W=V$ minimize the distance
$D=\text{min} _{W \in U(d^2)} \|U^R-W\|^2_{HS}$  where $\|X\|_{HS}=\sqrt{\tr(XX^{\dagger})}$ is the Hilbert-Schmidt norm. A direct variation of $D$ with $W$ subject to 
the unitarity constraint shows that this is related to the polar decomposition. It is known that if $U^R=V \sqrt{U^{R\, \dagger} U^R}$ is the polar decomposition of the general matrix $U^R$,  then the unitary operator $V$ is closest to it. In fact, it is the closest under any unitarily invariant norm (such that for any pair of unitary matrices $u$ and $v$, $\|uXv\|=\|X\|$) \cite{Fan1955,Keller1975,marshall2009inequalities}.

Given any unitary $U_0$, we find $U_n =M^n_R(U_0)$.  We show below that the distance of $U_n^R$ to it's nearest unitary is non-increasing with iteration $n$. Also observe that dual unitary operators are fixed points of the $M_R^2$ map.  Hence it seems plausible that if these fixed points are attracting, $U_n$ tends to become dual unitary. In particular, that $E(U_n)$ increases with $n$ and ideally towards the maximum possible value of $E(S)=1-1/d^2$. We found overwhelming numerical evidence for this, as shown in Fig.~(\ref{fig:EUvsn}), where we start from typical representatives from the CUE and the increase is not only monotonic but remarkably it is asymptotic to the maximum possible $E(U)$, thus  getting arbitrarily close to dual ones. For qubits, $d=2$, the approach is exponential: $\Delta_n =E(U)_{\text{max}}-E(U_n) \sim e^{-\alpha n}$.
For qutrits, $d=3$, the convergence of $E(U)$ to the maximum value of $8/9$ is also exponential
as shown in Fig.~\ref{fig:EUvsn}. The rates $\alpha$ are observed to be strongly dependent on the seed unitary $U_0$.
See Supplementary Material \cite{SM} for the cases of larger dimensionality $d$.
The approach for $d>4$ seems to be a power law on the average, as we explore large $n$. We found rare exceptions, only for the case of $d=3$, when $U_0$ was selected as some permutations the final operator reached was not dual.

From extensive numerical evidence, we conjecture that under the $M_R$ map, almost all unitaries sampled according to the CUE
monotonically tend arbitrarily close to being dual. Additionally, we are able to prove that
for any local dimension $d$, the $q=1/2$ Tsallis entropy of operator entanglement $S_{1/2}(U_n)=2(d \tr \sqrt{U_n^{R\, \dagger} U_n^R }-1)$, 
a monotonic function of the trace-norm $\|U_n^R\|_1=\tr\sqrt{U_n^{R\, \dagger} U_n^R}$, is non-decreasing under the $M_R$ map: 
\beq
\label{eq:Tsallishalf}
S_{1/2}(U_{n+1}) \geq S_{1/2}(U_n).
\eeq
This is an immediate consequence of the
the trace-norm $\|U_n^R\|_1$ itself being non-decreasing under the $M_R$ map: $ \|U_{n+1}^R\|_1 \geq  \|U_{n}^R\|_1.$ 

To prove this, let 
 \beq
 \label{eq:Dnsq0}
  D_n^2 =\text{min} _{W \in U(d^2)} \|U^R_n-W\|^2_{HS} =\|U_n^R-U_{n+1}\|_{HS}^2
  \eeq
 as $U_{n+1}$ is the nearest unitary to $U^R_n$.
From the observations that (i) the realignment is involutive, that is $(X^R)^R=X$, and (ii) the Hilbert-Schmidt norm is invariant $\|X^R \|_{HS}=\|X\|_{HS}=\sqrt{\sum_{ij} |X_{ij}|^2}$ under realignment as it is simply a permutation of the matrix elements, it follows that 
\beq
\label{eq:Dnsq}
D_n^2=\text{min} _{W \in U(d^2)} \|U_n-W^R\|^2_{HS} =\|U_n-U_{n+1}^R\|_{HS}^2.
\eeq
By definition,
\[ D_{n+1}^2=\text{min} _{V \in U(d^2)}\|U_{n+1}^R-V\|^2_{HS}, \]
hence using the Eq.~(\ref{eq:Dnsq}), it follows that $D_{n+1}^2\leq D_n^2$. From Eq.~(\ref{eq:Dnsq0}) we get 
\[
D_n^2= 2 d^2- 2 \text{Re}\tr(U_{n+1}^{\dagger} U_n^R)=2 d^2- 2 \tr(\sqrt{U_n^{R\, \dagger} U_n^R }).
\]
Hence the trace-norm of $U_n^R$ is a non-decreasing  function of $n$. As the corresponding extensive R\'enyi entropy $2 \log(\tr \sqrt{U_n^{R\, \dagger} U_n^R }/d)$ is also a monotonic function of the trace-norm, it inherits the same property of being non-decreasing under the introduced map.

Numerical evidence also points to the increase of any of the Tsallis (or R\'enyi) entropies, indicating that 
a majorization of the kind $U_{n+1}^{R\, \dagger} U_{n+1}^R \prec U_n^{R\, \dagger} U_n^R$ generically holds.   For two Hermitian operators $X$ and $Y$, $X$ is said to be majorized by $Y$ denoted as $X \prec Y $, if and only if $X = \sum_i p_i U_i Y U_i^\dagger$. Where $U_i's$ are unitary matrices and $0\leq p_i \leq 1$, $\sum_i p_i =1 $ \cite{alberti1982stochasticity}. Note that $X \prec Y$ imply  $S_q(X) \geq S_q (Y), \forall q >0$ \cite{ nielsen2001majorization, marshall2009inequalities}.

\emph{Characterization of dual unitaries via their entangling power}: An important characterization of the dual unitaries created is given by the other invariant $E(US)$ \cite{Zanardi2001,Bhargavi2019} ($S$ being the {\sc swap} operator), or equivalently its entangling power. The connection of the operator entanglements of $U$ and $US$ introduced above to entanglement in the bipartite space of states is via the entangling power.  This quantity is independent of any special property
of the initial unentangled state, as it is  the average linear entropy produced while operating on product states sampled uniformly. 
Let $|\psi_{AB}\kt = U|\psi_A \kt |\psi_B\kt$, $\rho_A=\tr_B(|\psi_{AB}\kt \br \psi_{AB}|)$, and $\mathcal{E}(\rho_A)=1-\tr(\rho_A^2)$ the linear entropy, then the entangling power $e_p(U)= \br \mathcal{E}(\rho_A)\kt$, where the averaging over an ensemble of $|\psi_{A}\kt |\psi_B\kt$.  If these are chosen from the Haar measure, then $e_p(U)=d^2[E(U)+E(US)-E(S)]/(d+1)^2$ \cite{Zanardi2001}.  Note that for dual unitaries, as $E(U)=E(S)$, $e_p(U)$ is simply proportional to $E(US)$, thus the entangling power is the main characterizer of this set.
 The maximum possible entangling power admitted by the dimensionality of the  spaces is $e_p^{\text{max}}=(d-1)/(d+1)$ \cite{Zanardi2000},  and is achieved by 2-unitary matrices for which $E(U)=E(US)=E(S)$.

 

\begin{figure}[htbp]
\centering
\includegraphics[height=.45\textwidth]{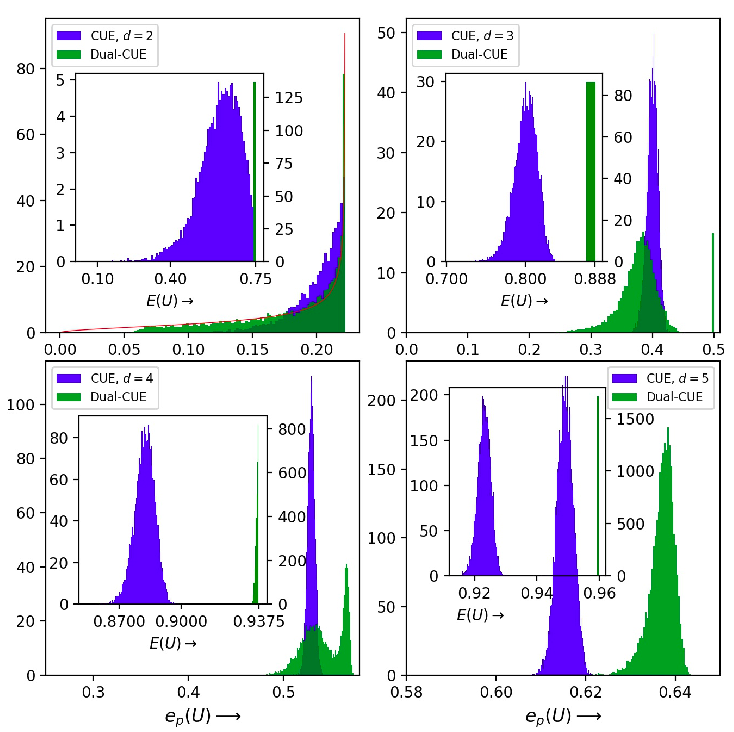}
\caption{The distribution of the entangling power $e_p(U)$  for both the CUE ensemble and the one obtained by iterating it with the $M_R$ map $n$ times, indicated as the ``dual-CUE" ensemble. The inset shows the distribution of the operator entanglements $E(U)$, notice the different scales on the vertical axes, left for the CUE and right for the dual-CUE. Here $n=10,10,30,50$ for $d=2,3,4,5$ respectively, and 10,000 realizations from the CUE are chosen in each case for the seed unitaries. }
\label{fig:EUSdistr}
\end{figure}

We construct an ensemble of unitary matrices starting from the CUE and iterating them under the $M_R$ map. Symbolically this ensemble is $M_R^{\infty}(\text{CUE})$, which we will refer to simply as ``dual-CUE", although in practice of course we will iterate a finite number of times to find intermediate ensembles. Figure~(\ref{fig:EUSdistr}) insets show the distribution of $E(U_n)$ for some appropriate choice of $n$ for the dual-CUE and for comparison the distribution of $E(U)$ for the CUE is also shown. The dual-CUE's entanglement seems to be tending to a Dirac delta function at the maximum value of $E(S)=1-1/d^2$ justifying the adjective, and this happens at smaller number of iterations $n$ for smaller $d$. 

In the main part of the figure is shown the distribution of the entangling power for the CUE and the dual-CUE. 
For small dimensions, the entangling power of the dual-CUE is broader and the mean of the entangling power is actually less than that of the CUE. For $d=2$, there is a divergence of the distribution corresponding to the dual-CUE around the maximum entangling power of $2/9$ (less than $1/3$ allowed by the dimensions and same as that of the {\sc cnot} and {\sc dcnot}  \cite{Collins2001} gates). 

In the case of qubits, the nonlocal part of the Cartan form of two-qubit gates is $\exp(-i c_1 \sigma_x \otimes \sigma_x-i c_2 \sigma_y \otimes \sigma_y-i c_3 \sigma_z \otimes \sigma_z)$ \cite{Zhang2003}, where $0\leq c_3 \leq c_2 \leq c_1 \leq \pi/4$ 
lies within a half of the tetrahedral  Weyl chamber \cite{Mandarino2018}, and we identify gates with $\pm c_3$.
 It is possible to derive a map of the parameter $c_i$ induced by the $M_R$ map \cite{ASL2020}, and indeed there is a fast convergence of these to $c_1=c_2=\pi/4$, which characterize dual unitaries. The 
final value of $c_3$ depends on the initial unitary. 
\begin{figure}[htbp]
\includegraphics[width=.45\textwidth]{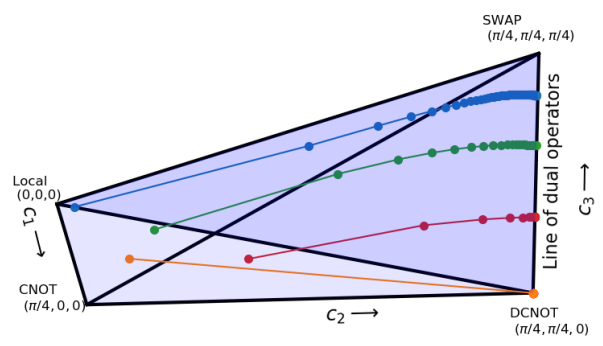}
\label{fig:cartan}
\caption{Trajectories under the nonlinear $M_R$ map in half of the Weyl chamber of 4 representative initial unitaries as they limit to the edge filled with dual unitary operators. All operators starting from the base reach the dual {\sc dcnot} gate in one iteration. }
\end{figure}
Such parametrization of  dual unitaries for qubits has been pointed out recently \cite{Bertini2019,Bhargavi2019}.

For the case of qutrits, the dual-CUE distribution is split and there is a peak at the largest possible value of $1/2$. In these cases, remarkably, the map $M_R$ has driven random CUE realizations into  2-unitaries   which maximize not just $E(U)$ but also $E(US)$ and hence the entangling power.
Approximately about $6\%$ of the CUE end up being of this kind. For the case of $d=4$, there is still a bimodal distribution, but the peak has shifted away from the maximum possible value of $3/5$, while for $d>4$, the distribution is not bimodal but also the dual-CUE distribution is more entangling than the CUE and the average entangling power is larger now for the dual-CUE. 

\noindent {\sl Generating 2-unitaries:} 
One may maximize $E(US)$ instead of $E(U)$ and this involves replacing realignment with partial transpose as $E(US)=1-\tr[(U^{T_A} U^{T_A \, \dagger})^2]/d^4$, where  $\br m \alpha | U^{T_A}| n \beta \kt= \br n \alpha |U| m \beta \kt$  ,  and we seek unitaries $U$ such that its partial transpose $U^{T_A}$ is also unitary. In place of the $M_R$ map, there is now a  partial transpose based one denoted $M_T$. 
We do not know if there exists a (dream) map that result in 2-unitaries or perfect tensors maximizing the entangling power which is a sum of $E(U)$ and $E(US)$, or equivalently $U$, $U^R$ and $U^{T_{A}}$ are all unitary. However, we found that the iteration of the composition  $M_{TR}$  wherein $R$ map is followed by the $T$ before finding the nearest unitary, often results in such operators for $d=3$ and $d=4$. 
\begin{center}
\begin{figure}[htbp]
\includegraphics[height=0.325\textwidth]{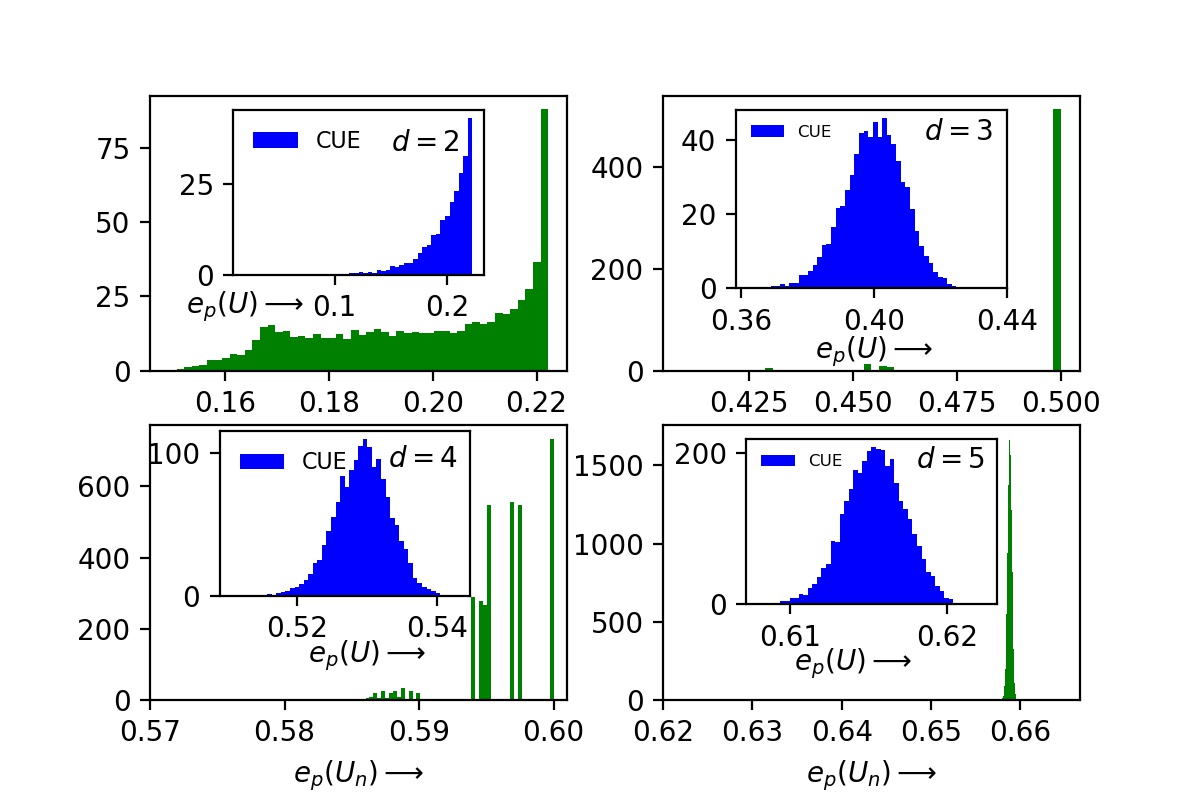}

\caption{The distribution of entangling power $e_p(U_n)$ under the $M_{TR}$ map, after $n=2000$ iterations and $10000$ realizations. For the cases of $d=3,4$ 2-unitaries or perfect tensors achieving the maximum value of $e_p$ are realized. The insets show the distributions of $e_p(U)$ for the corresponding CUE for comparisons.}
\label{fig:ep2unitdistr}
\end{figure}
\end{center}
This is shown in Fig.~(\ref{fig:ep2unitdistr}) where we see that while for $d=2$, the map produces a broad distribution of entangling powers as there are no 2-unitaries in this case. For $d=3$ we see a large peak at the maximum value of $1/2$. Remarkably, more than $95\%$ of the CUE seeded matrices end up being 2-unitaries, this may be contrasted with just the $M_R$ map, see Fig.~(\ref{fig:EUSdistr}) which led to about $6\%$. The remaining ones increase the 
entangling power but asymptote to lower values, as we also see for the case of $d=4$, where we have a complex set of prominent values. But unlike just the $M_R$ map which did not produce any 2-unitaries, about $20\%$ of those iterated with the $M_{TR}$ map end up being for all practical purposes, 2-unitaries.
For $d=5$, there is only one peak seen, but unlike $d=3,4$ these do not seem to get arbitrarily close to 
2-unitaries, and instead asymptote to about $98.9\%$ of the maximum allowed value. This leads to the disappointment that this map does not shed any light on the open problem concerning the existence of 2-unitaries in $d=6$ \cite{GWAZ2020,Open2020}.
The reason the $M_R$ map is ineffective in creating 2-unitaries for $d>3$  could be that they are much more constrained than dual unitaries and are required to be common fixed points \emph{both} of $M_R^2$ and $M_T^2$ maps.

To characterize the 2-unitaries obtained, we have run out of invariants such as $E(U)$, $E(US)$ or $e_p(U)$ as they are all maximized.
The distribution of the entanglement in states obtained by applying the 2-unitary operator to Haar distributed product states is the same for all locally equivalent operators. Remarkably, results not displayed indicate that such distributions are statistically indistinguishable from each other for all 2-unitaries obtained via the above procedure and also identical to those obtained from other known 2-unitaries such as permutations and Hadamard matrices\cite{Clarisse2005,Goyeneche2015}. Thus this leaves open the question of how different are the nonlocal contents of the 2-unitaries obtained thus far.

As far as the usual RMT properties such as the nearest neighbor spacing distribution and the form factor \cite{Mehta,Haake}, are concerned we have checked that the CUE is numerically indistinguishable from that of the dual-CUE or even the ensemble of 2-unitaries. The only way these map 
driven ensembles are different seem to be their nonlocal properties and their entangling abilities. 

\emph{Summary and open problems}: We have introduced nonlinear maps in the space of bipartite unitary operators whose fixed points are 
attracting and have generically maximal entangling properties. These produce, starting from the CUE, an ensemble of dual unitaries for any local dimensions and an ensemble of  2-unitaries  for local dimension $3$ and $4$, in turn producing  a large class of four partite AME states of qutrits and ququads. Many questions concerning the attractors and basins of attractions of these maps, which are novel dynamical systems in their own right remain open. Many-body systems built out of such special unitaries  could further reveal relations between entanglement, complexity and the nature of dynamical evolution. Other open problems include the extension of the current studies to multipartite systems \cite{linowski2019}, to unequal dimensional subsystems, and finding corresponding perfect tensors if they exist.

\begin{acknowledgments}
It is a pleasure for AL to thank Karol {\.Z}yczkowski and Zbigniew Pucha\l{}a for discussions on these matters over many years.
\end{acknowledgments}

%

\pagebreak

\newcounter{equationSM}
\newcounter{figureSM}
\newcounter{tableSM}
\stepcounter{equationSM}
\setcounter{equation}{0}
\setcounter{figure}{0}
\setcounter{table}{0}
\makeatletter
\renewcommand{\theequation}{\textsc{sm}-\arabic{equation}}
\renewcommand{\thefigure}{\textsc{sm}-\arabic{figure}}
\renewcommand{\thetable}{\textsc{sm}-\arabic{table}}

\begin{center}
{\large{\bf Supplemental Material for\\
 "Creating ensembles of dual unitary and maximally entangling quantum evolutions"}}
\end{center}

\section{Correspondence between states and operators}
In this section we summarize well-known aspects of the correspondence between operators and states \cite{Zyczkowski2004}, 
which is essentially the Choi-Jamialkowski isomorphism \cite{choi1975completely,jamiolkowski1972linear}. 
Linear operators acting on the states in  $d$ dimensional Hilbert space $\mathcal{H}_A^d$ itself form a $d^2$ dimensional Hilbert space endowed with 
Hilbert-Schmidt norm. Every operator $X \in \mathcal{H}_A^d$ can be isomorphically mapped to a state $\ket{X} \in \mathcal{H}_A^d \otimes \mathcal{H}_{A'}^d$ in a bipartite Hilbert space as, 
\begin{equation}
\begin{split}
\ket{X} &= \sum_{ij} X_{ij}|ij\kt \equiv \frac{1}{\sqrt{d}} \sum_{ij} \br i |X|j \kt |i \kt |j\kt=\frac{1}{\sqrt{d}} \sum_j X|jj\kt\\
=&(X \otimes I) \ket{\Phi^+}_{AA'},\quad \ket{\Phi^+}_{AA'} := \frac{1}{\sqrt{d}}\sum_i^d \ket{ii},
\end{split}
	\label{eq:iso}
\end{equation}
where we have defined the operator $X$ via its matrix elements $\br i |X|j\kt =\sqrt{d} X_{ij}$ and $\{\ket{i}\}_1^d$ forms an orthonormal basis in $\mathcal{H}^d$.

\begin{figure}[h]
\begin{tikzpicture}[scale = 0.7]
\fill[blue,rotate=90,fill opacity=0.1] (-1,2) ellipse (5 and 2.5);
\path [thick,draw=black!60!green,snake it]
    (-2,3) -- (-2,-5);
\node[thick] at (-2,3){\Large{$\bullet$}};
\node[thick] at (-2,3.5){$A'$};
\node[thick] at (-2,-5){\Large{$\bullet$}};
\node[thick] at (-2,-5.5){$A$};
\fill[red,rotate=128,fill opacity=0.1] (-1.5,-0.5) ellipse (6.2 and 2.5);
\path [thick,draw=black!60!green,snake it]
    (4.5,3) -- (4.5,-5);
\draw[rectangle] (-2.5,-5.8) -- (-1.5,-5.8) -- (-1.5,-4.5) -- (-2.5,-4.5) -- cycle; 
\draw[rectangle] (4,-5.8) -- (5,-5.8) -- (5,-4.5) -- (4,-4.5) -- cycle; 
\node[thick] at (4.5,3){\Large{$\bullet$}};
\node[thick] at (4.5,3.5){$B'$};
\node[thick] at (4.5,-5){\Large{$\bullet$}};
\node[thick] at (4.5,-5.5){$B$};
\draw (-2,-6.5) -- (4.5,-6.5);
\draw (-2,-5.8) -- (-2,-6.5);
\draw (4.5,-5.8) -- (4.5,-6.5);
\node[thick] at (1.2,-7) {\Large{$U$}};
\node[thick,blue,rotate=270] at (-2.8,-1.2) {$E(U)=1-\text{tr}(\rho_{AA'})^2$};
\node[thick,blue,rotate=268] at (-3.5,-1.2) {$\rho_{AA'}=\frac{1}{d^2}U^RU^{R^\dagger}$};
\node[thick,red,rotate=310] at (2.5,-1.5) {$E(US)=1-\text{tr}(\rho_{A'B})^2$};
\node[thick,red,rotate=308] at (2,-1.8) {$\rho_{A'B}=\frac{1}{d^2}U^{T_A}U^{T_A\dagger}$};
\end{tikzpicture}
\caption{$U$ acting on Hilbert space $\mathcal{H}^d_A \otimes \mathcal{H}^d_B$ is isomorphic to state $\ket{U}_{AA^\prime BB^\prime}$. Association of $E(U)$ and $E(US)$ with reduced states $\rho_{A^\prime A}$ and $\rho_{BB^\prime}$ is shown.   }
\label{fig:4party}
\end{figure}
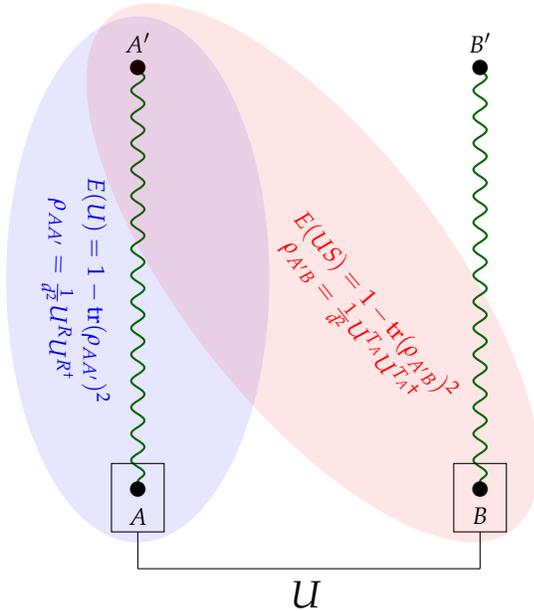

If the operator $U_{AB}$ is itself bipartite and defined over a Hilbert space $\mathcal{H}^d_A \otimes \mathcal{H}^d_B$ then the corresponding state is a four party state in $\mathcal{H}^d_A \otimes \mathcal{H}^d_{A^\prime} \otimes \mathcal{H}^d_B \otimes \mathcal{H}^d_{B^\prime}$ as shown in Fig. (\ref{fig:4party}). The four party state $\ket{U}_{A A' B B^\prime}$ is given by 
\begin{equation}
\ket{U}_{A A' B B^\prime} = (U_{AB} \otimes I_{A'B'}) \ket{\Phi^+}_{A A'} \ket{\Phi^+}_{BB^\prime}.
\end{equation}
The entanglement in the bipartition $AB|A'B'$ is maximal as $U_{AB}$ is unitary and local to the bipartition, and $|\Phi^+\kt$ are maximally entangled states. The entanglement in the bipartition $AA'|BB'$ is determined by the reduced density matrix $\rho_{ A A^\prime} = \frac{1}{d^2} U^R U^{R\dagger}$,
where $U^R$ is the realignment of $U$. Thus this is maximal iff $U^R$ is unitary. In general the entanglement in this bipartition is
the {\sl operator entanglement} as $U^R$ determines the operator Schmidt decomposition of $U$. In particular the operator
entanglement $E(U)$ used in the main text is the linear entropy of entanglement of this bipartition: $E(U) = 1 - \tr(\rho_{A A'}^2)$

Similarly, the entanglement in the bipartition $A'B|AB'$ is determined by the partial transpose, say with respect to subsystem $A$: $U^{T_A}$
as $\rho_{A^\prime B} = \frac{1}{d^2} U^{T_A}U^{T_A \, \dagger}$. The entanglement in this partition is the other local unitary invariant
operator entanglement:
$E(US) = 1 -\tr(\rho_{A^\prime B}^2) $, and is maximum iff $U^{T_A}$ is unitary, equivalently $\rho_{A^\prime B}$.
It is then evident that the entangling power $e_p(U) = d^2 [E(U)+E(US)-E(S)]/(d+1)^2$ \cite{Zanardi2000} is maximum if and only if {\sl both} $E(U)$ and $E(US)$ are maximum. As the three bipartitions discussed above are all that is required for determining the entanglement of any other
bipartition, the four party state is an absolutely maximally entangled one iff $e_p(U)$ is maximum.

\section{Growth of linear entropy $E(U)$ for higher dimensions ($d>3$):}
\begin{figure}
\includegraphics[height=.45\textwidth]{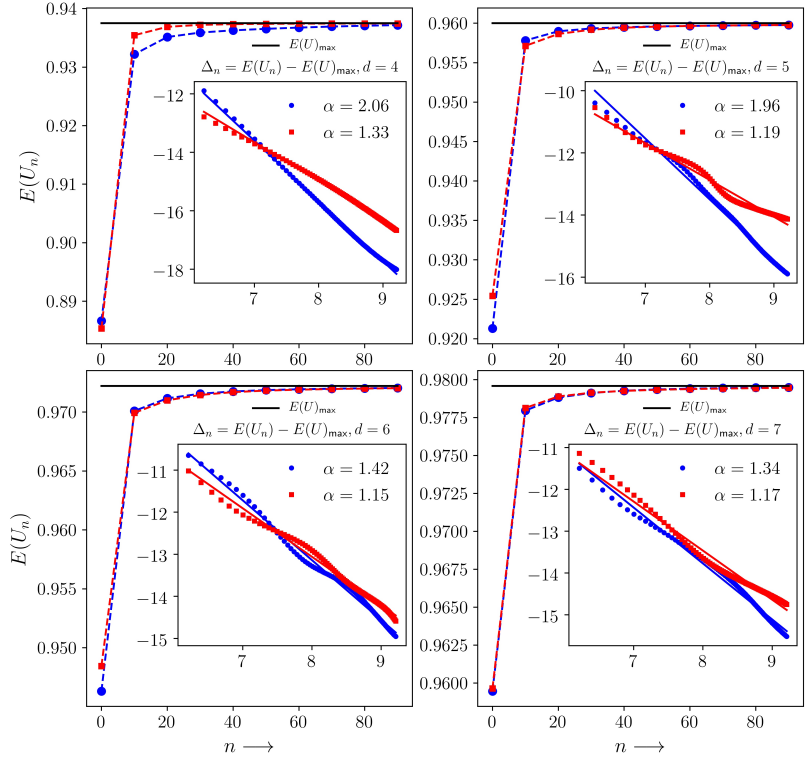}
\caption{Growth of $E(U_n)$ for $d=4,5,6,$ and $7$. $E(U_n)$ (shown for $n=0$ to $n=100$ in steps of $10$) saturates to the maximum possible value $(d^2-1)/d^2$ for sufficiently large $n$ for two random realizations. Shown in the insets are the corresponding log$(\Delta_n$)-log($n$) plots in the limit of large $n$ ($500 \leq n \leq 10,000$).}
\label{fig:EUhigherd}
\end{figure}

Under iteration of the $\mathcal{M}_R$-map, the linear entropy operator entanglement $E(U_n=\mathcal{M}_R^n U_0)$ of a randomly picked unitary $U_0\, \in \mathcal{U}(d^2)$ approaches the maximum possible value $(d^2-1)/d^2$ \cite{Zanardi2001} monotonically. The approach to the respective maxima for a $4 \leq d\leq 7$ is shown in Fig.~(\ref{fig:EUhigherd}).
The difference, $\Delta_n=E(U)_{\text{max}}-E(U_n)$, decays as a power law ($\Delta_n\propto n^{-\alpha}$) in the limit of large $n$ ($500 \leq n\leq 10,000$) with the exponent $\alpha$ depending on the seed unitary and dimensionality $d$.

\section{Distribution of linear entropy $E(U)$ under $\mathcal{M}_R$-map:}
Here we give some additional data about the distribution of the operator entanglement enroute to becoming a dual unitary.
An ensemble of randomly picked seed unitaries $U_0$ iterated $U_n=\mathcal{M}_R^n[U_0]$ yield an ensemble whose operator entanglement 
gets sharply peaked at the maximum possible value $=E(S)=(d^2-1)/d^2$ as $n \rightarrow \infty$. The intermediate $E(U_n)$ distributions for a large number of random realizations as shown below for different dimensions.

For qubits, $d=2$, due to the exponential saturation of $E(U_n)$ under the $\mathcal{M}_R$-map all random realisations are very close to the maximum possible value of $3/4=0.75$ for even for a relatively small value of $n\approx20$. For higher dimensions the distributions are seen to become
more sharply peaked, piling up at the maximum value, $E(S)=1-1/d^2$. Please note that the scale is different for each subplot in Fig.(3) to Fig.(5).

\begin{figure}[h]
\includegraphics[height=.45\textwidth]{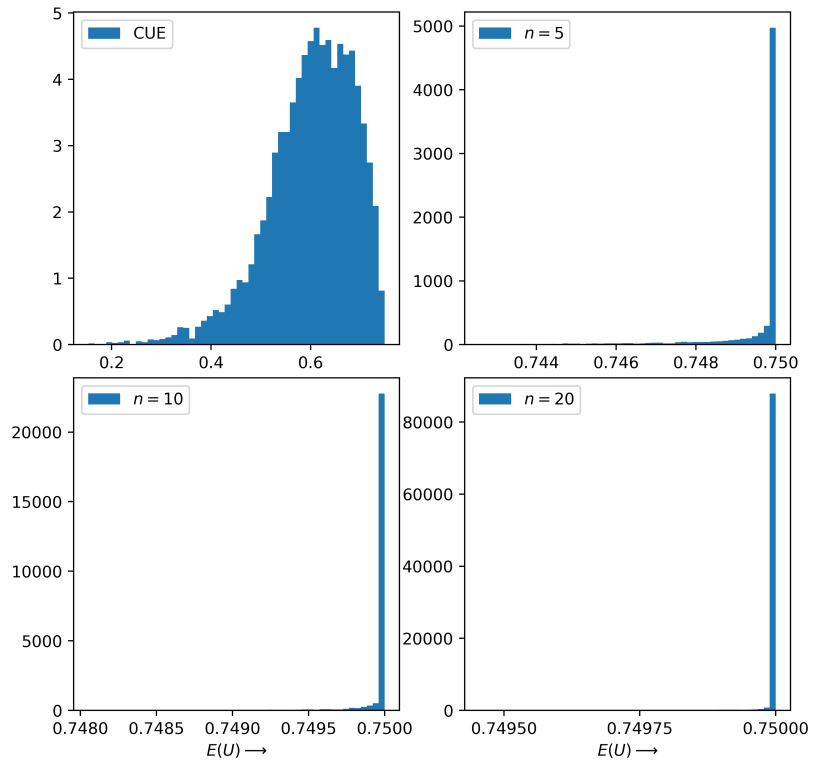}
\caption{Distribution of $E(U_n$) for CUE and $n=5,10,20$ under $\mathcal{M}_R$-map for $d=2$. For $n=20$, all randomly picked unitaries (around 10,000 in number) have reached almost close to the maximum possible value $3/4=0.75$.}
\label{fig:EUdistd2}
\end{figure}

\begin{figure}[h]
\includegraphics[height=.45\textwidth]{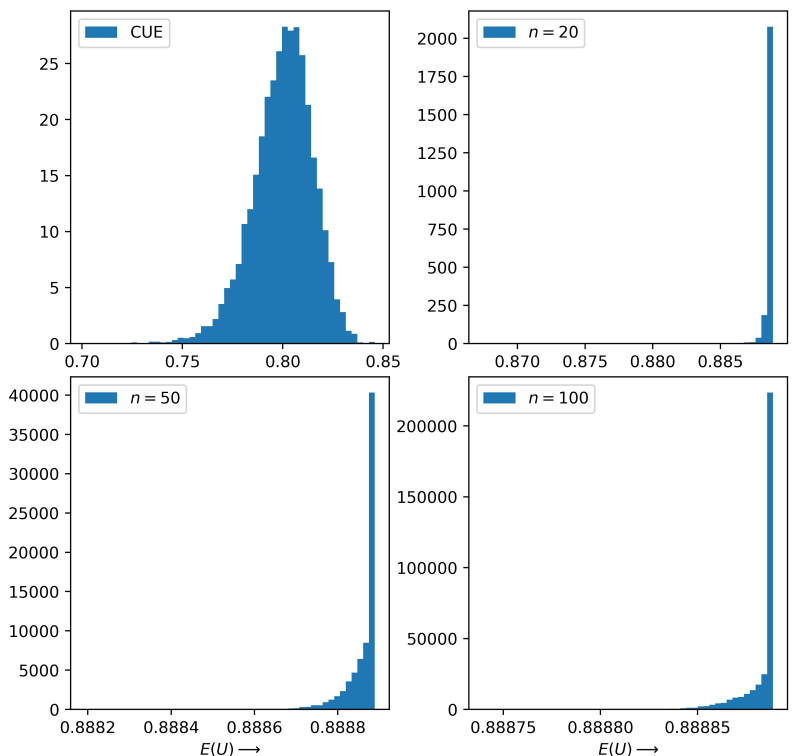}
\caption{Distribution of $E(U_n)$ under $\mathcal{M}_R$-map for $d=3$. Maximum possible value is $8/9=0.\overline{88}$.}
\label{fig:EUdistd3}
\end{figure}

\begin{figure}[h]
\includegraphics[height=.45\textwidth]{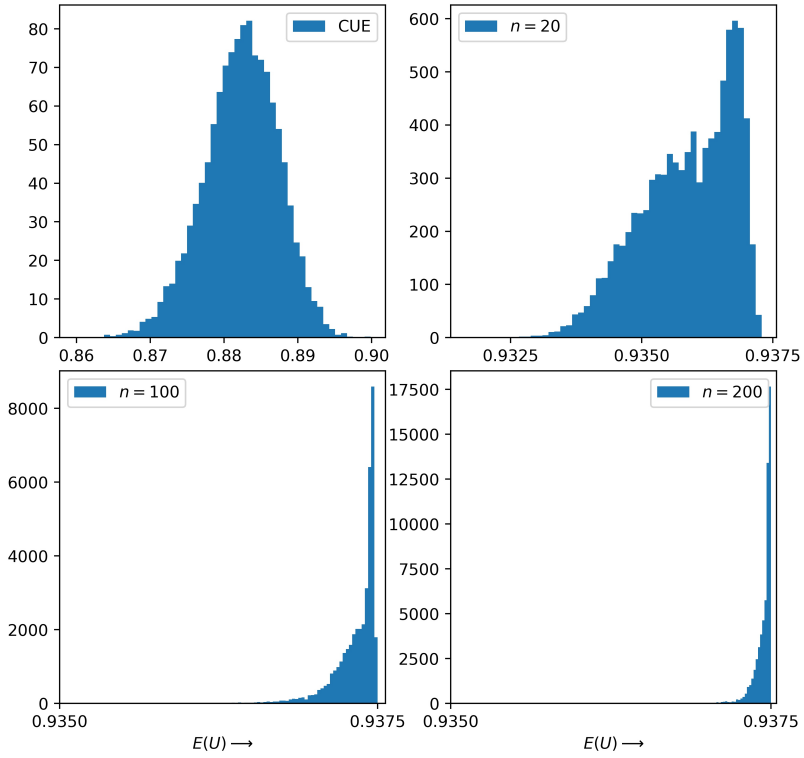}
\caption{Distribution of $E(U_n)$ under $\mathcal{M}_R$-map for $d=4$. maximum possible is $15/16=0.9375$ }
\label{fig:EUdistd4}
\end{figure}

\end{document}